\documentclass[preprint2]{aastex}

\begin{document} 

\lefthead{Graff and Kim}
\righthead{Using SNAP for Microlensing}
\title{Using the Supernova / Acceleration Probe (SNAP) to Search for
Microlensing Events Towards the LMC}
\author{David S. Graff}
\affil{University of Michigan, Department of Astronomy, Ann Arbor, MI 48109}
\authoremail{dgraff@astro.lsa.umich.edu}
\author{Alex Kim\altaffilmark{1}}
\altaffiltext{1}{Lawrence Berkeley National Laboratory, Berkeley, CA 94720}
\affil{University of California, Department of Physics, Berkeley, CA 94720}
\authoremail{agkim@lbl.gov}

\def\nth{n_{\rm th}}
\def\nobs{n_{\rm obs}}
\def\dmin{d_{\rm min}}
\def\macho{{\sc macho}}
\def\newpage{\vfill\eject}
\def\vs{\vskip 0.2truein}
\def\gnu{\Gamma_\nu}
\def\fnu {{\cal F_\nu}}
\def\mass{m}
\def\lum{{\cal L}}
\def\imf{\xi(\mass)}
\def\ilf{\psi(M)}
\def\msun{M_\odot}
\def\zsun{Z_\odot}
\def\met{[M/H]}
\def\vi{(V-I)}
\def\mtot{M_{\rm tot}}
\def\mhalo{M_{\rm halo}}
\def\pp{\parshape 2 0.0truecm 16.25truecm 2truecm 14.25truecm}
\def\la{\mathrel{\mathpalette\fun <}}
\def\ga{\mathrel{\mathpalette\fun >}}
\def\fun#1#2{\lower3.6pt\vbox{\baselineskip0pt\lineskip.9pt
  \ialign{$\mathsurround=0pt#1\hfil##\hfil$\crcr#2\crcr\sim\crcr}}}
\def\ie{{ i.e., }}
\def\eg{{ e.g., }}
\def\etal{{et al.\ }}
\def\etalc{{et al., }}
\def\kpc{{\rm kpc}}
 \def\Mpc{{\rm Mpc}}
\def\mh{\mass_{\rm H}}
\def\mmax{\mass_{\rm u}}
\def\ml{\mass_{\rm l}}
\def\bc{f_{\rm cmpct}}
\def\br{f_{\rm rd}}
\def\ibl{{\cal I}(b,l)}
\def\dmax{d_{\rm max}}
\def\dmin{d_{\rm min}}
\def\mbol{M_{\rm bol}}
\def\kms{{\rm km}\,{\rm s}^{-1}}

\begin{abstract}
Microlensing experiments today have a tantalizing result; they have
detected an excess of microlensing events beyond what is expected from
known stellar populations.  These events could be due to a possible
form of Halo dark matter.  However, study of these events is limited
by their small number and poor photometric precision.  The SNAP
satellite, which has been proposed to search for supernovae, would be
an ideal instrument to search for microlensing events towards the LMC.
Due in part to its larger mirror, and largely to its superb
space-based seeing and low background, with a modest program SNAP
will be able to detect $\sim 250$ microlensing events per year.  In
addition, SNAP will generate $\sim 16$ high quality events per year with
1\% photometric resolution.  These data should break the microlensing
degeneracy, and determine whether the events are indeed due to dark
matter lenses in the Galactic Halo.
Side benefits of such a program will include a never before equaled
catalog of variable stars in the LMC.  On of the most intriguing possibilities is a direct measurement of the distance to the LMC.  Such a measurement will determine what is today the most uncertain rung on the distance ladder and thus lock down the Hubble constant.

\end{abstract}


\section{Introduction}

The principal motive for creating microlensing searches was to search
for dark matter in the form of discreet lumps, or MACHOs, by searching
for microlensing towards the Magellanic Clouds.  A major result of
these experiments has been to show that the dark matter is not
composed of objects with mass in the range $10^{-7} - 10^{-1} \msun$
(Alcock \etal 1998, Lasserre \etal 2000).  The MACHO group has claimed
an intriguing positive detection of lensing events significantly above
that expected from known stellar populations.  However, these events
are difficult to interpret owing to the degeneracies inherent in
microlensing: the sole observable in an event, its timescale, is a
function of three parameters, the lens mass, its parallax, and its
proper motion.

If the MACHO events are interpreted as being due to lenses which are part of
the halo, then roughly 20\% of the mass of the halo has to be made of
MACHOs with mass about $0.5\msun$.  This result would be quite
significant, representing the first detection of a component of the
Milky Way dark matter.  If the MACHOs were baryonic, their mass
suggests that they might be white dwarfs, an interpretation which has
been strengthened by a possible detection of halo white dwarfs by
Ibata \etal (1999) although their Hubble Deep Field proper motion survey
is not consistent with larger ground-based proper motion surveys
(Flynn \etal 2000).  However, white dwarfs make poor dark matter
candidates (Graff \etal 1999; Fields, Freese \& Graff 1998, 2000), so
there is a real chance that the microlensing experiments have detected
a new non-baryonic dark matter candidate such as primordial black
holes (Carr \& Hawking  1974; Jedamzik 1997).

Microlensing's ability to detect Halo dark matter will always
be cast in doubt as long as the microlensing events could be due to a
possible background.  One possible background is a rare population of
variable stars, hitherto unknown because their frequency is so low
$(\sim 10^{-8})$.  Some of the detected microlensing events have high
signal-to-noise and so precisely follow the microlensing light curve
that their interpretation is unambiguous.  Other microlensing events
are detected with less signal-to-noise.  During the course of the
microlensing experiments, a whole class of such stars was discovered
and eliminated from the samples, the so called ``blue bumpers''.
However, there may be other sorts of variable stars lurking in the H-R
diagram.  For example, the EROS group has shown that one star in the
LMC had a bump in its light curve resembling sufficiently a microlensing
curve to be classified as a microlensing event.  Five years
later, its light curve had a similar excursion, suggesting that at
least this star is a variable.

Another possible background is a previously unknown population of
ordinary stars acting as either sources or lenses, sometimes called
``self lensing''.  These stars would cause true microlensing, but
would not indicate the presence of dark matter.  Although severe
constraints have been put on this hypothesis (Gyuk, Dalal \& Griest
2000), it is still viable.  Weinberg (2000) has suggested that the LMC 
could be distorted out of virial equilibrium by the tidal forcing of 
the Milky Way in such a manner as to have a large thickness (and thus
a large microlensing optical depth) while still maintaining the low
velocity dispersion measured by several groups, and reviewed by Gyuk,
Dalal \& Griest (2000).  Graff \etal (2000) report a possible
detection of such a population along the line of sight to the LMC.
For a recent review of the interpretation of microlensing events, see
Graff (2000).

In sum, microlensing experiments may have detected a possibly
non-baryonic form of Dark Matter, making up a good fraction of the
mass of the Milky Way halo.  However, this claim cannot be verified
until all possible background sources of events, such as previously unknown
populations of variable stars and lensing due to ordinary stars have
been eliminated.

A variety of ways have been proposed to break this degeneracy, which
we shall review later in this paper.  These techniques either
require very high signal to noise events, to look for subtle
deviations from the standard microlensing light curve, or they require many
more events than have been detected; either to do meaningful
statistical tests on the events, or to look for rare exotic events.

We will show that the proposed SNAP (Supernova / Acceleration Probe)
satellite
would be an ideal tool to search for microlensing towards
the Large Magellanic Cloud, and perhaps towards the Small Magellanic
Cloud.  With a relatively small amount of observing time, it could
detect orders of magnitude more events than the current ground-based
experiments.  In addition, it would generate a reasonable ensemble of
high signal-to-noise events, which would eliminate all doubts that
these events are indeed microlensing events, and allow them to be
classified as Halo lenses or LMC lenses.

\section{SNAP}

Both microlensing searches and supernovae searches require similar
telescopes that are used in similar manners since both experiments
involve searching for a point-like object that gets brighter and then
dims over the course of tens of days.  In fact, the MACHO group has
found detected supernovae to be a substantial background in their
experiment (Alcock \etal 2000a) and their telescope has been used to
search for supernovae (Reiss \etal 1998),
while the EROS group has a supernova search program
using their telescope (Hardin \etal 2000).  In both types of
experiment, a large field is sampled frequently, with a frequency of
hours to weeks.  The ``new'' image is subtracted from a reference
field, and one looks for a point-like bright spot, which could be
either due to a supernova, or the amplification of a microlensed star.
In both types of experiments, a ``trigger'' allows detection of the
event in real time, allowing for further, more intensive study.  The
primary difference between the two types of experiments is that
microlensing observations must be done towards fields rich in source
stars with low background contamination, while supernova searches are
best done towards deep galaxy-rich fields with few foreground objects.

The Supernova / Acceleration Probe (SNAP) is a proposed satellite-borne
telescope designed specifically for the detection and high-precision
observation of cosmological supernovae.  Since searching for
microlensing events is so similar to searching for supernovae, it is
unsurprising that both the hardware and software capabilities of the
SNAP program are ideally suited to microlensing.
SNAP's mission objectives require a large field imager,
high signal-to-noise observations of point sources, frequent
light-curve sampling, and suffer from telemetry and pointing
constraints.  In the following, what we describe as ``SNAP'' is based on
a feasibility-study design
(available at \url{http://snap.lbl.gov}).
Ongoing design and trade studies
will formalize (and perhaps modify)
SNAP's mission parameters.  They should not, however, do
so in a way
that significantly changes the
characteristics important for microlensing science.

SNAP's wide field of view is provided by
a one square-degree optical imager.  It and
the optical telescope assembly's  photon gathering
capabilities are distilled concisely into Table~\ref{exptime},
which gives the signal-to-noise obtained from 140 second
exposures of point sources of differing magnitude.  Note that the
noise is dominated by the source itself and not background zodiacal light.

\begin{deluxetable}{c|cccccc}
\tablewidth{0pt}
\tablecolumns{7}
\tablecaption{Signal to Noise for exposure of 140 sec\label{exptime}}
\tablehead{&\multicolumn{6}{c}{Magnitude}\\
Filter & 20 &	21 &	22 & 	23 &	24 &	25}
\startdata
V &	135 &	85 &	53 &	32 &	19 &	10 \\
R &	157 &	99 &	61 &	37 &	22 &	12 \\
I &	128 &	80 &	49 &	29 &	16 &	8 \\


\enddata
\label{texposetable}
\end{deluxetable}

SNAP's current design maintains a fixed single face directed
towards the Sun; this constrains observing fields with full-time
accessibility to be towards the
ecliptic poles.  Fortunately, however, the LMC is nearly at the South
Ecliptic pole, with an ecliptic latitude of $-85^\circ$.  The SMC would be a
bit more difficult, with an ecliptic latitude of $-65^\circ$.  The
Galactic center could not be observed by SNAP under its current design
since it is near the ecliptic.

A modest search for $t_E \sim 20-100$
day LMC microlensing events could be folded into or incorporated in
the supernova search.  Finer temporal resolution could be achieved with a
more ambitious microlensing survey using more
of SNAP's observing resources.

\section{SNAP's ability to detect microlensing events}

In this section, we will estimate how many events could be detected by
a microlensing program involving SNAP.  We will follow the optimal
microlensing formalism developed by Gould (1999).  This formalism
assumes that the microlensing experiment will use image subtraction to
find microlensing events even when the source star is not isolated in
the field, sometimes known as ``Pixel Lensing''.  SNAP will be well
suited for pixel lensing since the image subtraction technique will
already be implemented for its supernova search program.

The number of microlensing events detected will depend on the strategy
adopted by the microlensing experiment.  For the purpose of this
paper, we shall assume that SNAP takes images with exposures of 140
seconds (which can accommodate its current feasibility-study draft
telemetry rate).  Following this
strategy, SNAP could image 120 deg$^2$ of the LMC in 5.3 hours.  The
bulk of observations should be made in one filter.  Occasional
observations can be be made in other pass-bands to provide color
information on variable stars.

SNAP fields will be directed
towards both Ecliptic poles and revisited with a frequency of $\sim 4$
days.  We will assume that while SNAP is pointed towards the South
Ecliptic pole it will observe the entire LMC in addition to its
supernova fields.  This sampling frequency of four days is short
compared to the typical measured Einstein Radius crossing time of 40
days and the shortest measured Einstein Radius crossing time of 18
days, so these events should be detected without much trouble.
Additional target of opportunity time could be allocated for deeper,
better time-sampled photometry and color measurements of particularly
interesting microlensing events.

Following the Gould (1999) estimate, the number of microlensing events
that can be detected with a particular $\chi^2$ significance level
depends on the signal-to-noise ratio of individual observations, and
on the number of stars in the field, which in turn depends on the
luminosity function of the LMC and the surface brightness of the field
in question.  The signal-to-noise of the current generation of ground
based experiments is severely limited by the crowding of several stars
into their super-arcsecond seeing.  With its larger mirror, and most
importantly, with its superb space-based seeing, SNAP should easily
achieve a signal-to-noise far in excess of MACHO or EROS, or even the
proposed next generation experiment (Stubbs 1998).  Furthermore,
unencumbered by weather and the moon and covering every one of the
120 fields during each observing run, its efficiency, the fraction of
microlensing events occurring which are actually detected, should be
close to 100\% as compared with less than 50\% for the MACHO experiment
(Alcock \etal 2000a) and less than 30\% for the EROS experiment
(Lasserre \etal 2000).


We estimate the expected number of microlensing events to be found
by our proposed
experiment by scaling from the ground-based
``next generation'' microlensing experiment proposed
by Stubbs (1998) and whose detection rates were given in Figure~2
of Gould (1999).  The majority of SNAP events will come close to the
detection threshold (e.g. $R=24.5$ for a conservative $S/N=15$) where
a 140 second SNAP exposure is equivalent to
a 220 minute ground exposure at new moon\footnote{At these magnitudes,
SNAP is source noise dominated whereas the
ground experiment is sky dominated;
SNAP is thus faster by the ratio of sky (168 e$^-$/sec)
and source fluxes (3 e$^-$/sec).  Mirror apertures
and quantum efficiencies give similar counts for the same source.}.
From Gould (1999) this observing depth will produce
$\sim 245$ events per year
of current ground-based quality across the entire face of the LMC, factoring
in the difference between theoretical and real MACHO events.

The ground based experiments frequently see microlensing in unresolved
source stars.  They interpret these events by using HST to count how
many unresolved stars there are.  SNAP too will measure microlensing
in unresolved source stars, but HST followup will also be unable to
resolve these stars.  Interpretation of these SNAP events will require
either modeling the LMC luminosity function or measuring the
luminosity function with a better resolution telescope such as NGST.

In addition to the shear increase in quantity, SNAP will also provide
a large sample of high quality events.  An examination of Table
\ref{texposetable} shows that under the campaign we propose, SNAP will
have 1\% photometric resolution on all events brighter than $V<21$.
Nearly all LMC events discovered to date by the ground-based
experiments are at least this bright, thus, all of these events would
be followed with 1\% photometric resolution.  Since SNAP would cover a
larger fraction of the LMC than the ground based experiments under our
test observation strategy, and since its efficiency would be higher,
it would find roughly 4 times as many bright events as the MACHO
collaboration finds (one factor of 2 coming from the higher
efficiency, and one factor coming from the increased area of our
proposed SNAP survey), or perhaps 10 bright events per year.

We can make a quantitative estimate of this number as follows:  The
detection rate of microlensing events is
\begin{equation}
\Gamma=\frac{2}{\pi}N_*\frac{\tau \, \epsilon \, u_0}{t_E}
\end{equation}
where $N_*$ is the number of observed stars, $\tau$ is the optical
depth, $\epsilon$ is the efficiency of detection, $u_0 \approx 1$ is
the minimum impact parameter for an event to be classified a
``detection'', and $t_E \approx 40$ days is the typical Einstein
radius crossing time of the event.  This method is applicable in the
case where the source stars are not crowded, true for the brighter
sample of stars under consideration here with the expected SNAP
resolution.  According to the MACHO collaboration (Alcock \etal 2000a),
the optical depth towards the LMC is roughly $\tau \approx 10^{-7}$.
Using the luminosity function of Alcock \etal (2000b), we
see that there are roughly $N_*\approx 2.5 \times 10^7$ stars brighter
than $V<21$ in the LMC.  Assuming that the efficiency of detecting
such bright events is $\epsilon \approx 1$, we expect an event rate of
16 bright events per year.  

\section{How will SNAP data resolve the Microlensing Degeneracy?}

There are several proposed techniques to determine whether the
observed lensing is indeed due to lenses in the Halo.  These
techniques have heretofore been limited by the low  quantity and
quality of microlensing events.  However, the two orders of magnitude
increase in quantity of SNAP events, and order of magnitude increase
in photometric precision will allow these techniques to come into
their own.

As noted above, SNAP should detect some 16 events with 1\% photometric
resolution and occasional color measurements.
If these events conform to microlensing lightcurves, there
should be no further doubt that they are due to microlensing and not
to some exotic unclassified variable star.

Gould (1998) suggested that if a typical LMC microlensing event is
measured every day with 1\% photometric resolution, then one could
measure the leading term of its microlensing parallax (the
mean acceleration of the earth) (Gould 1992), a
small deviation from the standard microlensing light curve due to the
motion of the earth around the sun.  In that case, one can still not
completely break the degeneracy of the microlensing event, but it
should still be possible to separate halo lensing (which should have
detectable parallax) from LMC lensing (which should not).  This
technique could only be applied to events brighter than $V<21$ and  as
discussed above, SNAP should see $\sim 16$ of these events per year.

To measure the parallax of these 16 events, they should be followed up
daily.  Since only 4 of these events will be visible at any one time
on average, this daily follow-up requires an additional 18 minutes per
day on average.  This follow-up need not be done with SNAP with its
enormous field of view: with proper coordination, HST could serve
instead.

The microlensing parallax signal scales as the square of the event
time scale, so we would only be able to measure the parallax of the
longer events.  Thus, the parallax technique would not be sensitive to
the possibility that, for some reason, the longer events are due to
LMC self lensing while the shorter events are due to, e.g., brown
dwarfs in the halo.  Therefore, we note other methods SNAP will be
able to use to break the microlensing degeneracy which will make use of
the shear quantity of microlensing events detected by SNAP.

If the microlensing events are truly due to Halo lensing, then we
would expect only a small change in the optical depth and time scales
of microlensing events across the face of the LMC.  The MACHO group
(Alcock \etal 2000) has already used this technique to rule out one
particular model of LMC self lensing, but has not ruled out all
realistic models of self lensing, being hampered by the relatively low
number of events in their sample, and by the limited spatial extent of
their fields.  The SNAP program, with its ability to detect hundreds
of events across the face of the LMC, would be able to clearly rule
out models of LMC self lensing.

In addition to being uniform across the face of the LMC, the optical
depth must also be uniform across the H-R diagram.  Zhao, Graff \&
Guhathakurta (2000) have suggested that under LMC self lensing models,
the source stars will tend to be towards the back of the LMC.  They
should thus should be somewhat more reddened and somewhat dimmer than
the LMC field stars.  The MACHO group has looked for this effect in
their sample, but has been unable to draw firm conclusions owing to
the relatively small number of events in their sample (Alcock \etal
2000c).  Also, by verifying that the much larger sample of
events are evenly drawn from the HR diagram, SNAP will confirm that
the events are true microlensing events and not due to some heretofore
unknown type of variable star.

By generating a large number of events, SNAP will generate a decent
sample of rare events, such as finite source size events (Alcock \etal
1997), amplification of a binary source (Han \& Gould 1997) or events
due to a binary lens (Schneider \& Weiss 1986).
All of these variations on the standard microlensing curve allow a
partial breaking of the microlensing degeneracy, by uncovering one of
the two hidden parameters, the parallax of the lens and its proper
motion (relative to the source in both cases).  Still, this partial
breaking will be enough to simply distinguish between the Halo-lensing
hypothesis and the self-lensing hypothesis, though it won't be able to
do finer measurements of, for example, the mass function of the lenses.

  As shown by Honma (1999), high amplification events and binary lens
events should be completely solvable if intensely monitored from a
satellite such as SNAP by using the motion of the satellite as
its orbit moves the satellite in and out of the caustic.  The non-linear
motion of the satellite as the satellite orbits the earth causes a small (1\%)
deviation in the shape of the light curve during the caustic crossing.
This technique will only work if the caustic crossing time scale is
about as long, or longer than the satellite orbit time, 7--14 days.
Thus, the Honma technique can only be applied to events which
are known to be self lensing events, which have much longer caustic
crossing time scales than halo lensing events.  Still, this technique,
by directly solving the event, will resolve the mass function of the
binary population of the LMC.  That is, we will be sure of the mass of
the components of the lens, whereas now, we have to estimate the mass
of the lens statistically, and based on assumptions of the parallax
and proper motions of the lenses.

In summary, SNAP will be able to break the microlensing degeneracy
using several independent techniques, and thus unambiguously determine
whether or not the lenses lie in the Halo or in the LMC.  SNAP thus
has the potential to identify a large component of the dark matter in
the Galactic Halo.

\section{Non-microlensing programs}

In addition to the microlensing results from this LMC program, there will
be other scientific results which may be just as interesting as the
microlensing results themselves.  It is beyond the scope of this work
to detail them all, and we challenge the astronomical community,
especially the experts in variable stars, to
propose further benefits of these observations.

\subsection{Variable stars}

Microlensing surveys have already begun a revolution in the study of
variable stars (e.g. Bauer \etal 1999, Alcock \etal 1999, Udalski
\etal 1999).  The two order of magnitude increase in depth and
precision made possible by SNAP cannot help but increase the pace of
this revolution.

The ground based microlensing experiments have been limited in their
studies to relatively bright, high amplitude stars, such as Cepheids.
SNAP will be able to generate similar catalogs of dimmer variable
stars, such as $\delta$ Scuti stars, and of stars with low amplitudes,
such as CP stars (Hensberge 1993).  SNAP will also possibly discover
new types of variable stars which had been missed in the past because
they are rare and have small amplitudes.

There are too many types of variable stars to detail individually in
this paper.  However, we can lay out the general benefits of such a
program that will apply to all types of variable stars.

All the variable stars found by SNAP will be at the same distance (to
the extent that the LMC is thin), and with relatively low reddening,
facilitating the construction of color magnitude diagrams of
variables, perhaps yielding their mass, age, and metallicity.  The
efficiency of finding variable stars will be constant across the face
of the LMC and well modeled; thus we will be able to characterize the
frequency of variable stars.  The high resolution of SNAP will find
many variable stars in LMC clusters, allowing a measurement of their
age.  Perhaps most importantly, the LMC has a lower metallicity than
the Milky Way, though the age and perhaps environments are similar, so
studies of LMC variable stars will be able to examine the effect of
metallicity on variable stars.

For all variables which are driven through internal pulsations, even
those with low amplitude, SNAP will generate precise light curves.
Studies of the detailed shapes of light curves of stars driven by
internal pulsations will yield new measures of the internal structure
of these stars.

\subsection{Distance to the LMC}

The distance to the LMC is at present the most uncertain rung in the
cosmic distance ladder.  Cepheids in the LMC are well studied (and
will be even better studied by SNAP as discussed in the previous
section) and are used to calibrate the distance to other galaxies,
which may in turn be used to calibrate other distance indicators.
With a major campaign involving the Hubble Telescope, much of the
traditional uncertainty surrounding the Hubble Constant has
dissipated, to the extent that the reported values of the Hubble
constant tend to vary within a narrower band than the factor of two
common ten years ago.  For example, Gibson et al. (2000) report a
value of $H_0=68 \pm 2 \pm 4 \kms {\rm Mpc}^{-1}$ while Saha et
al. report a value of $H_0 = 60 \pm 2 \kms {\rm Mpc}^{-1}$.

Both of these measurements assume that the distance to the LMC is 50
kpc, but the present day uncertainty of the distance to the LMC is
larger than the reported uncertainties (which do not include the LMC
distance), or than the difference between these two values.  In
contrast to the Hubble Constant, there continues to be disagreement
over the distance to the LMC.  This distance is usually measured with
standard candles, lately normalized against local candles calibrated
with Hipparcos.  For example, Feast and Catchpole (1997) find a
distance to the LMC of $55\pm 2.5$ kpc using trigonometric parallaxes
of Cepheids while Udalski (2000), using a variety of distance indicators,
finds a value of 44 kpc. 

By co-adding 100 images of the LMC, SNAP should in effect generate a
total exposure of 14,000 seconds covering the entire LMC.  This will
generate an HR diagram containing 100 million stars measured with 1\%
photometric resolution, down to $V<25$, or down to absolute magnitude
of $M_V\sim 6.5$.  This data set will allow a larger group of standard
candles to be used to calibrate the LMC distance.  The extra precision
will also allow a study of the orientation and three dimensional
structure of the LMC.

The most satisfying solution would be to directly measure the distance
to the LMC.  With such a detailed study of the
LMC, and with such high resolution, it may be possible to directly
measure the distance via trigonometric parallax.  The ability of SNAP
to measure parallax will be limited by its relative astrometric
precision.  For the moment, in this paper, we will discuss the purely
statistical limits (due to photon noise, and a limited number of
astrometric sources) to the astrometric precision.

The astrometric precision, especially in the case where the seeing is
of order the pixel size (the $V$ band is undersampled for SNAP) will
depend on the details of the point spread function and the detailed
shape of the individual pixels.  Given the possibility of dithering
and the stability and planned study of the PSF, we can assume that the
astrometric precision is roughly the seeing times the photometric
uncertainty.  Thus, for a single pointing towards a 20th mag star,
SNAP should have an statistical astrometric precision of $\sim 1.5
mas$.

The more crowded LMC fields may have some $10^6$ stars as bright as
magnitude 20, and we are interested in the average parallax of these
stars.  Hence, the astrometric precision may be limited by the number
of distant astrometric references rather than by the number of
foreground LMC stars.  There are about 30 quasars brighter than mag 20
in each 1 degree square SNAP field, not enough to achieve the required
statistical precision.  Ibata et al. (1999) used compact bright
galaxies as their reference.  They found 50 in the 0.0015 deg$^2$
Hubble field of view, so there should be $\sim 3 \times 10^5$ in the
SNAP field (roughly the same as the number of brighter stars).

With an exposure every 4 days for a period of 4 years, there will be
365 total exposures.  Thus, the total statistical parallax precision
towards the LMC is roughly
\begin{equation}
\sigma_\pi \approx 0.14\, \mu as \left ( \frac{365}{N_{\rm exp}} \frac{3 \times 10^5}{N_{\rm obj}} \right )^{1/2} \, .
\end{equation}
This estimate of the parallax precision is about 0.6\% of the parallax of the LMC, $20 \mu as$, compared to the present day uncertainty of about 20\%.

This estimate only deals with the purely statistical uncertainties
involved with such a program.  This project will obviously be limited
by systematic uncertainties, which cannot be estimated at this stage
in the design of the satellite.  These systematic errors may prove too
large to be useful.  Nonetheless, with such high statistical
precision, it is worth pursuing such a program in the hopes that the
systematic uncertainties can be beaten down to an interesting level.

\section{Conclusion}

We have shown that the SNAP satellite will have orders of magnitude
greater sensitivity to microlensing events than ground-based searches.
A relatively modest program, one which can be done concurrently with
the primary mission of SNAP to search for supernovae, can cover the
entire LMC and find $\sim 245$ events per year.  This program should also
find some 16 events per year with 1\% photometric resolution.
Armed with these data, it will be possible to identify the location of
the lenses; are they in the galactic Halo or in the LMC?  


Any microlensing program would also have several important side
benefits, including a new variable star study of the LMC, and the
construction of an extremely deep compendium of LMC stars.  One novel
side benefit may be a new precise direct measurement of the distance
to the LMC, allowing a more accurate determination of the Hubble
constant and extragalactic distance scale.

Part of this work has been funded by NSF Grant 21434-13066 and by the
University of Michigan.  We thank Andy Gould, Saul Perlmutter, Scott Gaudi, and
Alain Milsztajn for careful readings of the manuscript.

\end{document}